\documentclass[aps, prd, twocolumn, lengthcheck, superscriptaddress, showpacs, letterpaper, nofootinbib]{revtex4-1}

\usepackage{epsfig}
\usepackage[usenames]{color}
\usepackage{graphicx}
\usepackage{amsmath}
\usepackage{epstopdf}

\newcommand\sect[1]{\emph{#1}---}
\def\bi{\bibitem}

\def\la{\langle}\def\ra{\rangle}
\def\be{\begin{eqnarray}}\def\ee{\end{eqnarray}}
\def\lsim{\mathrel{\rlap{\lower3pt\hbox{\hskip1pt$\sim$}}
     \raise1pt\hbox{$<$}}} 
\def\gsim{\mathrel{\rlap{\lower3pt\hbox{\hskip1pt$\sim$}}
     \raise1pt\hbox{$>$}}} 
\def\del{\partial}

\allowdisplaybreaks

\begin{document}

\title{Nuclear Axial Currents from Scale-Chiral Effective Field Theory}

\author{Yan-Ling Li}
\affiliation{Center for Theoretical Physics and College of Physics, Jilin University, Changchun,
130012, China}

\author{Yong-Liang Ma}
\affiliation{Center for Theoretical Physics and College of Physics, Jilin University, Changchun,
130012, China}

\author{Mannque Rho}
\affiliation{Institut de Physique Th\'eorique,
	CEA Saclay, 91191 Gif-sur-Yvette c\'edex, France }
\date{\today}

\begin{abstract}
By incorporating hidden scale symmetry and hidden local symmetry in nuclear effective field theory, 	combined with double soft-pion theorem, we predict that the Gamow-Teller operator coming from the space component of the axial current should remain unaffected by the QCD vacuum change caused by baryonic density whereas the first forbidden beta transition operator coming from the time component should be strongly enhanced. While the latter has been confirmed since some time, the former is given a support by a powerful recent {\it ab initio} quantum Monte Carlo calculation in light nuclei, also confirming the old ``chiral filter hypothesis."  Formulated in terms of Fermi-liquid fixed point structure of strong-coupled nuclear interactions, we offer an extremely simple resolution to the long-standing puzzle of ``quenched $g_A$", $g_A^{\rm eff}\approx 1$~\cite{quenched},  in nuclear Gamow-Teller beta transitions, giant Gamow-Teller resonances and double beta decays.

\end{abstract}

\maketitle
\sect{\bf Introduction}\label{sec:Intro}
The  behavior of the axial-vector coupling constant $g_A$  in nuclear medium has a long history of puzzles encompassing nuclear physics, astrophysics and particle physics. In nuclear physics, there has been the mysterious $\sim 20\%$ quenching of $g_A$ in shell-model calculations of nuclear beta decay,  giant Gamow-Teller resonances and double beta decay.  In particle physics, there is the issue of partial restoration of chiral symmetry, an intrinsic property of the symmetry of QCD, and in astrophysics, a surprising role of first-forbidden beta decay in nucleosynthesis. Some of these issues are comprehensively reviewed in \cite{gA-review}.

In this Letter, we propose a simple resolution of the $g_A$ problem based on scale-invariant chiral effective field theory combined with ``chiral filter" mechanism anchored on the current-algebra soft-pion theorems\footnote{The soft-pion theorem that is exploited here is {\it trivially} encoded in modern chiral perturbation theory but seems to have a deep theoretical connection with infrared structure of gravity as well as gauge theories. This point will be briefly mentioned at the end of this paper.}. We will argue that the recent  {\it ab initio} quantum Monte Carlo calculation by Pastore {\it et al.}~\cite{monte carlo}  goes a long way in giving the support to our simple solution.

\sect{\bf Scale-Chiral Symmetry}
Among symmetries most relevant to nuclear dynamics, QCD has chiral symmetry which is broken explicitly by the quark mass and scale symmetry which is broken explicitly by the trace anomaly.

How chiral symmetry figures in nuclear effective field theory is now well understood in the guise of chiral perturbation theory and fairly well-established in modern development of {\it ab initio} approaches since the paper of Weinberg~\cite{weinberg}. Since the quark masses involved in nuclear dynamics are tiny compared with the chiral scale,  $\sim 1$ GeV, one can talk about the ``chiral limit" where one approximates by setting the quark mass equal to zero.  In doing the calculations, it makes sense to theoretically ``turn off" the chiral symmetry explicit breaking.  When chiral symmetry is  spontaneously broken, there emerge Nambu-Goldstone bosons and they are the well-known pions $\pi$.

In contrast, while QCD has classically scale invariance in the chiral limit,  quantum mechanically, there is an anomaly, called trace anomaly, which posits a scale, and hence the scale symmetry is explicitly broken. This anomaly is renormalization-group invariant, and hence cannot be ``turned off" as the quark mass can be, so it seems there is no sense in talking about scale-invariance as is done with chiral symmetry. The explicitly broken scale symmetry can also be broken spontaneously generating a Goldstone boson, intrinsically massive due to the trace anomaly, so a pseudo-Nambu-Goldstone boson, called dilaton.  Now there is a  subtlety with scale symmetry spontaneous breaking, because unlike chiral symmetry, scale symmetry cannot be broken spontaneously unless explicitly broken~\cite{Freund-Nambu}.  This raises a conundrum in introducing scale symmetry, i.e., scalar meson dilaton, in nuclear physics~\cite{conundrum}.

In nuclear physics, there is a dire need for a local scalar field of mass $\sim 600$ MeV. Such a scalar plays an important role for nuclear forces as well as Walecka-type mean field approaches, i.e., energy density functional, to nuclear matter popular in nuclear physics. There is in the particle data booklet a low-mass scalar $f_0(500)$, and an attractive possibility is to identify it as the dilaton. This is the proposal made by Crewther and Tunstall (CT)~\cite{CT}. A similar idea was proposed by the authors of \cite{GS}.

We will follow here the approach by CT. In CT, a presence of an infrared fixed point in the QCD $\beta (\alpha_s)$ function is postulated, i.e., $\beta(\alpha_{\rm IR})=0$. The difficulty is that whether such an IR fixed point is present in QCD in the vacuum is not known. There is no indication either for or against such an IR fixed point for the number of flavors $N_f$ less than $\sim 8$\footnote{There is lattice indication for the presence at large $N_f$ near what is called ``conformal window" which has a connection with Higgs physics.}. This issue cannot be settled at the moment as explained in detail in ~\cite{conundrum}. Here we will bypass that conundrum by assuming that although it may not make sense in the vacuum, one can consider approaching the vicinity of an IR fixed point in medium  and study fluctuations around -- and not on top of  --  the potential IR fixed point. This is the standpoint we take here. With the assumption so made, we will make certain predictions to be confronted with nature.

There are two predictions in particular that are relevant to nuclear physics. One is the prediction for the properties of compacts stars, involving highly dense matter. Reference \cite{PKLMR} addresses this issue. We will not go into this matter. The other is the $g_A$ problem that we are interested in here.

As Yamawaki has argued~\cite{yamawaki}, scale symmetry that we are considering is present -- though ``hidden" -- in linear sigma model. Starting with linear sigma model -- to which the Standard Higgs model belongs, it has been shown by dialing a parameter in the model that the linear sigma model can be driven to the familiar non-linear sigma model on which chiral perturbation theory is built,
\be
{\cal L}_{{\rm NL}\sigma} = \frac{f_\pi^2}{4} {\rm Tr} \left(\partial_\mu U \partial^\mu U^\dagger\right) +\cdots
\ee
or to a scale invariant model with dilaton coupled to nonlinear sigma field with a dilaton potential that breaks scale symmetry. It is the latter form that is relevant to us and we suggest that it is the baryonic density that drives the coupling. In the chiral limit, it has the form
\be
  {\cal L}_{\rm Scale\sigma} &=& {\cal L}_{\rm sinv} - V(\chi)\,\label{L-dilaton}
\ee
with
\be
{\cal L}_{\rm sinv} &=& \frac{1}{2} \left(\partial_\mu \chi \right)^2+ \frac{f_\pi^2}{4}\left(\frac{{\chi}}{f_\sigma}\right)^2\cdot {\rm Tr} \left(\partial_\mu U \partial^\mu U^\dagger\right) +\cdots. \label{s-inv}\label{invLag}
 \ee
 Here $U$ is the usual chiral field, which is a scale singlet and $\chi$ is the mass dimension 1 ``conformal compensator field" $\chi=f_\sigma e^{\sigma/f_\sigma}$ -- where $\sigma$ is the dilaton field -- that transforms as a singlet under chiral transformation and scale dimension 1 under scale transformation. The ellipsis stands for higher order terms. As given, the Lagrangian (\ref{s-inv}) is scale invariant and chiral invariant.   All scale symmetry breaking, both explicit and spontaneous, are put in the dilaton potential $V(\chi)$ which does not figure explicitly in the process we are concerned with.

 There is another hidden symmetry in chiral Lagrangian that plays equally important role as the scalar dilaton $\chi$ and it involves the vector mesons $\rho$ and $\omega$. The mass involved is comparable to that of the scalar $\sim 700$ MeV so they need to be incorporated together. In fact the $\omega$ is essential in Walecka-type relativistic mean field approach providing the necessary repulsion while the $\rho$ comes into the nuclear tensor forces with the sign opposite to the pion tensor. The symmetry associated with the vector mesons is local gauge symmetry and hence what is involved is hidden local symmetry (HLS)~\cite{HLS}. This symmetry can be easily implemented to the scale-symmetric Lagrangian by exploiting the redundancy present in the chiral field $U$ to make Lagrangian (\ref{invLag}) hidden local symmetric. The resulting Lagrangian is scale-symmetric HLS, sHLS for short.

 Finally there remains how to set up the power series of scale symmetry in conjunction with chiral symmetry, the  power counting of which is well established. Following the idea of CT, we consider expanding around the assumed IR fixed point at which the beta function is zero, $\beta (\alpha_{\rm IR})=0$,
 \be
 \beta(\alpha_s)=\delta\cdot \beta^\prime  +\cdots
 \ee
 where
 \be
 \delta &=& (\alpha_s-\alpha_{\rm IR})\\
\beta^\prime &=& \frac{\partial}{\partial\alpha_s}\beta(\alpha_s)|_{\alpha_s=\alpha_{\rm IR}}
  \ee
is the anomalous dimension of $G_{\mu\nu}^2$ with $G_{\mu\nu}$ being the gluon energy momentum tensor.
The power counting in power expansion in scale-chiral perturbation theory is then
\be
O(\partial^2)\sim O(p^2)\sim O(m_\pi^2)\sim O(\delta).
\ee
Here $\pi$ stands for the octet pseudo-scalar NG bosons. Note that $\beta^\prime$ signaling explicit scale symmetry breaking is $O(1)$ in the scale-chiral counting in contrast to chiral symmetry where the chiral symmetry explicit breaking quark mass is counted as $O(p^2)$.

The above counting rule has been recently implemented in deriving scale-chiral expansion incorporating both hidden symmetries and the detailed discussions  are given in ~\cite{LMR}.  The formalism is applied in \cite{LM-BR} to derive Brown-Rho scaling in medium which is valid in the density regime up to $\sim 2n_0$.\footnote{Beyond $\sim 2 n_0$ where the skyrmion-half-skyrmion topology change takes place, the intrinsic density dependence coming from the matching of the correlators of  the EFT and QCD at a matching scale, not operative at density $n\lsim 2n_0$, must be taken into account.}

The expressions of the Lagrangian beyond the leading order, namely, $O(p)$ in the baryon sector and $O(p^2)$ in the meson sector, given in \cite{LMR,LM-BR} are quite complicated involving a large number of unknown parameters. But to the leading order (LO) in scale symmetry, it is simple to reproduce what's given in \cite{LMR,LM-BR}. Suppose one has a Lagrangian ${\cal L}^{(m)} (\Phi)$ involving matter fields $\Phi$ (baryons $B$, mesons  $\rho$, $\omega$, $\pi$) and the total scale dimension in the Lagrangian density is $m\leq 4$. Then one makes the Lagrangian density to have scale dimension 4, so that the action is scale-invariant, by multiplying it with the conformal compensator field $\chi$
as
\be
\bar{{\cal L}}=\left(\frac{\chi}{f_\sigma}\right)^{4-m} {\cal L}^{(m)}.
\ee
Then the CT procedure is to write
\be
\bar{\cal L}\to \left( \kappa  +(1-\kappa)\left(\frac{\chi}{f_\sigma}\right)^{\beta^\prime}\right)\bar{\cal L}
\ee
 where $\kappa$ is an unknown constant. Now when the dilaton field is turned off by setting $\sigma=0$, the $\beta^\prime$ dependence disappears. It will give the usual chiral perturbation theory, HLS if vector fields are included. Then there will be no footprint of scale symmetry breaking in it.

 There are two ways that the CT Lagrangian reduces to the form of the hidden scale symmetric sHLS. One is that $\beta^\prime \ll 1$, that is, weak explicit scale symmetry breaking. This is somewhat in the similar situation as the kaon mass with the dilaton mass of the same size. As there, perturbation expansion in $\beta^\prime$ could make sense. Another possibility is that $\kappa\approx 1$. This seems to be more consistent with the notion that scale symmetry is hidden and in fact is favored in the treatment of compact stars~\cite{PKLMR} where scale invariance is considered as ``emergent" or un-hidden at high density. Expanded to higher orders in $\beta^\prime$, the physics may be quite different, but at this order,
the resulting Lagrangian, with the explicit scale symmetry breaking entirely in the dilaton potential, can have an analogy to the usual chiral Lagrangian where the explicit symmetry breaking is put entirely in the quark mass term. We shall call this leading order scale symmetry (or LOSS) Lagrangian.

\sect{\bf Nuclear Axial Currents}
For our consideration in nuclear processes, we can restrict ourselves to chiral $SU(2)\times SU(2)$. Reducing from three flavors for which scale-chiral EFT is formulated~\cite{LMR,LM-BR} to two flavors, one can extract the relevant part of the Lagrangian which is found to be extremely simple
\be
 {\cal L} &=&i\overline{N} \gamma^\mu \del_\mu N -\frac{\chi}{f_\chi}m_N \overline{N}N +g_A \overline{N}\gamma^\mu\gamma_5 \tau_a N{\cal A}_{\mu}^a+\cdots\nonumber\\
 \label{LAG}
 \ee
 where ${\cal A}_\mu$ is the external axial field.
  Note that while the kinetic energy term and particularly the nucleon coupling to the axial field are scale-invariant by themselves and hence do not couple to the conformal compensator field, the nucleon mass term is multiplied by it. Put in the nuclear matter background, the bare parameters of the Lagrangian will pick up the medium VeV. Thus in (\ref{LAG}) the nucleon mass parameter will scale in density, while, significantly, $g_A$ will remain {\it unscaled}
 \be
 m_N^\ast/m_N &=& \la\chi\ra^\ast/f_\sigma\equiv \Phi\label{Phiscaling}\\
 g_A^\ast/g_A &=& 1\label{gA}
 \ee
 where $f_\sigma$ is the medium-free VeV $\la\chi\ra_0$ and the $\ast$ represents the medium quantities. The first is one of the scaling relations given in \cite{BR91}. The dilaton condensate carries density dependence when the vacuum is warped by density. It is a part of ``induced density dependence (IDD)" inherited from QCD valid at $n\lsim 2n_0$~\cite{PKLMR}. The second is new and says that the Lorentz-invariant axial coupling constant {\it does not} scale in density. This result was already indicated in the Skyrme term of the Skyrme model~\cite{BR91} but what's given here is more directly linked to QCD symmetries. Combined with the ``chiral filtering mechanism" to be specified below,  this is the most important  point in the present note.

 Now in medium, Lorentz invariance is spontaneously broken, which means that the space component, $g_A^{\rm s}$ and the time component $g_A^{\rm t}$ could be different. Indeed writing out the space and time components of the nuclear axial current operators, one obtains
 \be
\vec{J}_A^{\pm} (\vec{x}) &=& g^{\rm s}_A \sum_i \tau_i^{\pm} \vec{\sigma}_i \delta(\vec{x}-\vec{x}_i),\label{GT}\\
 J_{5}^{0\pm} (\vec{x})&=&- g^{\rm t}_A \sum_i\tau_i^\pm  \vec{\sigma}_i \cdot (\vec{p}_i  - \vec{k}/2) /m_N \delta(\vec{x}-\vec{x}_i)\label{axialcharge}
 \ee
 where $\vec{p}$ is the initial momentum of the nucleon making the transition and $\vec{k}$ is the momentum carried by the axial current.  In writing (\ref{GT}) and (\ref{axialcharge}),  the nonrelativistic approximation is made for the nucleon. This approximation is valid not only near $n_0$ but also in the density regime $n\gsim n_{1/2}\sim 2n_0$. This is because the nucleon mass never decreases much after the parity-doubling sets in at $n\sim n_{1/2}$ at which $m_N^\ast\to m_0 \approx (0.6-0.9) m_N$~\cite{PKLMR}. It will be related to pion decay constants below.

 A simple calculation taking into account (\ref{Phiscaling}) and (\ref{gA}) gives
\be
g_A^{\rm s}=g_A, \ \ g_A^{\rm t}=g_A/\Phi\label{main}
\ee
with $\Phi$ given by (\ref{Phiscaling}).

\sect{\bf Chiral Filtering Effect}
To confront with nature, we need two ingredients: (1) accurate nuclear wave functions; (2) reliable nuclear weak currents. In a systematic EFT calculations, the two are to be treated on the same footing. In practice a full consistency is not feasible and neither is it necessary. To proceed, let us suppose that the wave functions are accurately calculable with an accurate potential. Now regarding the point (2), the soft-pion theorems figure crucially. In the scale-chiral counting with the scheme espoused in this paper (which is essentially equivalent to chiral counting~\cite{TSP}), taking the axial current to be a ``soft pion," there is a soft-pion exchange current that involves double soft-pions coupling to the nucleon dictated by the current algebras. This term was shown to be the most important exchange current contribution to axial current transitions in nuclei. This was shown first in 1978 using soft-pion theorems~\cite{KDR} and in 1991 using chiral perturbation theory~\cite{MR91}. The phenomenon was dubbed ``chiral filter hypothesis" since at the time the high-powered computational techniques currently developed were not available to check quantitatively the arguments believed to be reasonable. The prediction was that there would be (a) a huge meson exchange correction due to soft pions to the one-body charge operator (\ref{axialcharge}) governing first-forbidden beta transitions,  with higher chiral corrections strongly suppressed, and (b) the Gamow-Teller transitions controlled by the space component of the axial current will be given {\it entirely} by the leading order one-body operator (\ref{GT}), given that next corrections are estimated to come only at a much higher order.

The enhancement factor for the axial-charge operator is very simple to calculate. Involving the soft-pion exchange the ratio of the two-body over one-body matrix elements $R$ can be computed almost nuclear model independently. It comes out to be $R=0.5\pm 0.1$ ranging from  $A=12$ to $A=208$. An extremely simple calculation shows that with the two-body effect taken into account,  the effective axial-charge operator is obtained by making the replacement in (\ref{axialcharge}) by~\cite{KR}
$g_A^t\rightarrow g_A^{t\ast} = \epsilon g_A$ with $\epsilon=\Phi^{-1} (1+R/\Phi)$.  The scaling factor $\Phi$ is related to the pion decay constant in medium $f_\pi^\ast$ as $\Phi\approx f_\pi^\ast/f_\pi$~\cite{PKLMR}. At nuclear matter density, one gets $\Phi(n_0)\approx 0.8$ from deeply bound pionic nuclei~\cite{yamazaki}. The enhancement factor at nuclear matter density is then $\epsilon (n_0)\approx 2.0\pm 0.2$, within the range of  theoretical uncertainty in the ratio $R$. This is confirmed by what was found in Pb nuclei~\cite{warburton}, $\epsilon^{\rm exp} (n_0)=2.01\pm 0.05$.  The results in $A=12, 16$~\cite{e-MEC} are compatible with the Pb result. It is an understatement to say that this is a gigantic correction as an exchange current effect.\footnote{Needless to say, {\it ab initio}  high-powered calculations on this matter in light nuclei would be highly desirable to further confirm this prediction.}

Now we turn to the other side of the coin of the chiral filter mechanism, i.e., the Gamow-Teller coupling constant. Here the soft-pions are rendered powerless and hence whatever corrections that come to the leading one-body operator must be suppressed relative to the leading $O(1)$ operator, accounting for  the ```chiral filtering." The quantum Monte Carlo calculations in $A=6-10$ nuclei by Pastore {\it et al} did verify  this feature at the N$^4$LO. At that order, the filter seems to work remarkably, say, at the level of $\lsim 3\%$. What is even more striking is that with high-order correlations accounted for in the Monte Carlo approach, there is no indication for $g_A$ quenching, which means that it is the high-order nuclear correlations in the wavefunctions, not a basic modification of the axial current, that could have been responsible for the ``$g_A$ problem."  If that is the case, then this calculation provides a support for the prediction that $g_A$ should not be affected by the chiral condensate decreasing with density as given by (\ref{main}). By the Goldberger-Treiman relation, this  means that the $g_{\pi NN}$ coupling should also be unaffected by the vacuum change. This is somewhat surprising. As is generally accepted, the pion decay constant should follow in some ways the chiral condensate which is considered to decrease as temperature or density increases, going to zero at chiral restoration. In the same vein, it has been considered plausible that the axial coupling constant would approach 1 as chiral symmetry is realized in Wigner-Weyl  mode, as is indicated at the dilaton-limit fixed point~\cite{PKLMR}. The prediction made in this paper and the powerful {\it ab initio} quantum Monte Carlo calculation indicate that this is not the case.

Given the above explanation of where the quenched $g_A$ is located, the question that remains is why is $g_A$ quenched ``universally" by $\sim 20\%$ in the nuclear shell model calculations~\cite{quenched}?

We offer an extremely simple answer in terms of Landau Fermi-liquid fixed point theory using the scale-chiral EFT Lagrangian, $s$HLS. The key ingredient for this is that the mean-field approximation with the $s$HLS Lagrangian {\it endowed with the IDDs inherited from QCD}  corresponds to Landau-Fermi liquid fixed point theory \`a la Wilsonian renormalization group to many-fermion systems with Fermi surface ~\cite{FR,PKLMR}. In the large $N$ limit where $N=k_F/(\Lambda-k_F)$ where $\Lambda$ is the cutoff on top of the Fermi surface,  the Landau mass $m_L$ and quasiparticle interactions $\cal {F}$ are at the fixed point, with $1/N$ corrections suppressed~\cite{shankar}. The relation between the Landau mass $m_L$ and the effective $g_A^{\rm L}$, both taken at the fixed point,  is given by~\cite{BR91,FR}
\be
\frac{m_L}{m_N}=1+\frac{1}{F_1}=\left(1-\frac{\tilde{F}_1}{3}\right)^{-1}\approx \Phi \sqrt{\frac{g_A^{\rm L}}{g_A}}
\ee
where $\tilde{F}_1$ is related to the Landau parameter $F_1$ by $\tilde{F}_1=(m_N/m_L)F_1$.   Applying the mean field argument, this relation gives
\be
\frac{g_A^{\rm L}}{g_A}\approx \left(1-\frac 13 \Phi\tilde{F}_1^\pi\right)^{-2}
\ee
where $\tilde{F}_1^\pi$ is the pion Fock term contribution to the Landau parameter $\tilde{F}_1$. The Fock term is a loop contribution, so naively $O(1/N)$. But the pion being ``soft," it plays an indispensable role as it does for the anomalous orbital gyromagnetic ratio $\delta g_l^p$~\cite{FR}.

Let us consider the $g_A^{\rm L}$ at nuclear matter density. With $\Phi(n_0)\approx 0.8$ inferred from deeply bound pionic systems~\cite{yamazaki} and $\frac 13 \tilde{F}_1^\pi (n_0) =-0.153$, which is precisely given by the pion exchange, we get, with the current value of $g_A=1.27$,
\be
g_A^{\rm L} (n_0) \approx 0.79 g_A\approx 1.0.\label{landaugA}
\ee
This is precisely $g_A^{\rm eff}$ needed in the shell-model calculations~\cite{quenched} and in the giant Gamow-Teller resonances~\cite{sakai}.  Note here the crucial role of the pionic contribution  interlocked with the dilaton condensate for the quenching. It turns out that the density dependence in $\Phi$ (dropping with density) nearly cancels the density dependence in
$\tilde{F}_1^{\pi}$ (increasing with density) so that the product  $\Phi\tilde{F}_1^\pi$ becomes independent of density\footnote{ The $g^L_A$ differs by less than 2\% between the densities $\frac 12 n_0$ and $n_0$.}. Thus the Landau $g_A$ (\ref{landaugA}), evaluated for nuclear matter,  applies not only to heavy nuclei but also to light nuclei.

Now how does this $g_A^{\rm L}$ correspond to $g_A^{\rm eff}$ in the shell-model calculations?

To answer this question, recall that {\it at the Fermi-liquid fixed point in our formulation,  the beta functions for the quasiparticle interactions $\cal{F}$, the mass $m_L$, the Gamow-Teller coupling $g_A^{\rm L}$ etc. at a given density  should be suppressed}. This means in particular that the loop corrections to the effective $g_A$ should be suppressed. It is therefore the effective coupling constant, duly implemented with density-dependent condensates inherited from QCD and with high-order quasiparticle correlations subsumed, to be applied to non-interacting quasiparticles, that is,  simple particle-hole configurations in shell model calculations.  This corresponds {\it effectively} to what is captured {\it microscopically} in the {\it ab initio} quantum Monte Carlo calculation of \cite{monte carlo}.

\sect{\bf ``Soft-Pion Theorem Triangle"}
The prominent role  soft pions play in the processes addressed above raises a potentially deep issue in nuclear physics. According to the lore of effective quantum field theory, it makes good sense to integrate out the pion for processes involving energy scales much lower than the pion mass, leading to what is known as  pionless ($\not\!\pi$)  EFT. It turns out such a $\not\!\!\pi$\,EFT works fairly well in a number of low-energy nuclear processes. Now the question would be how the soft-pion effect, crucial in certain nuclear processes such as first-forbidden beta transitions,  can manifest when pion fields do not figure explicitly? The interplay between the in-medium vacuum condensate $\Phi$ and the pionic Landau parameter $F_1^\pi$ is mysterious. This intriguing question may have an answer in the recent development involving soft theorems in the web of triangles ``echoed" in a variety of  infrared structure of gauge -- and gravity -- theories~\cite{strominger}. It may figure as a sort of ``memory effect" in the triangle with the soft-pion theorem sitting on one corner.

\subsection*{Acknowledgments}
Y.~L. Ma was supported in part by National Science
Foundation of China (NSFC) under Grant No. 11475071, 11547308 and the Seeds Funding of Jilin
University.

\end{document}